\documentclass[12pt,preprint]{aastex}

\slugcomment{}

\shorttitle{Finding Your Literature Match}
\shortauthors{Henneken et al.}

\begin{document}

\title{Finding Your Literature Match -- A Recommender System}

\author{Edwin A. Henneken, Michael J. Kurtz, Alberto Accomazzi, Carolyn Grant, Donna Thompson, Elizabeth Bohlen, Giovanni Di Milia, Jay Luker, Stephen S. Murray}
\affil{Smithsonian Astrophysical Observatory, 60 Garden Street, Cambridge MA 02138, USA}
\email{ehenneken@cfa.harvard.edu}

\begin{abstract}
The universe of potentially interesting, searchable literature is expanding continuously. Besides the normal expansion, there is an additional influx of literature because of interdisciplinary boundaries becoming more and more diffuse. Hence, the need for accurate, efficient and intelligent search tools is bigger than ever.

Even with a sophisticated search engine, looking for information can still result in overwhelming results. An overload of information has the intrinsic danger of scaring visitors away, and any organization, for-profit or not-for-profit, in the business of providing scholarly information wants to capture and keep the attention of its target audience. Publishers and search engine engineers alike will benefit from a service that is able to provide visitors with recommendations that closely meet their interests. Providing visitors with special deals, new options and highlights may be interesting to a certain degree, but what makes more sense (especially from a commercial point of view) than to let visitors do most of the work by the mere action of making choices? Hiring psychics is not an option, so a technological solution is needed to recommend items that a visitor is likely to be looking for. In this presentation we will introduce such a solution and argue that it is practically feasible to incorporate this approach into a useful addition to any information retrieval system with enough usage.
\end{abstract}


\section{Introduction}
\begin{figure} [hb]
\centering
\includegraphics[scale=0.75]{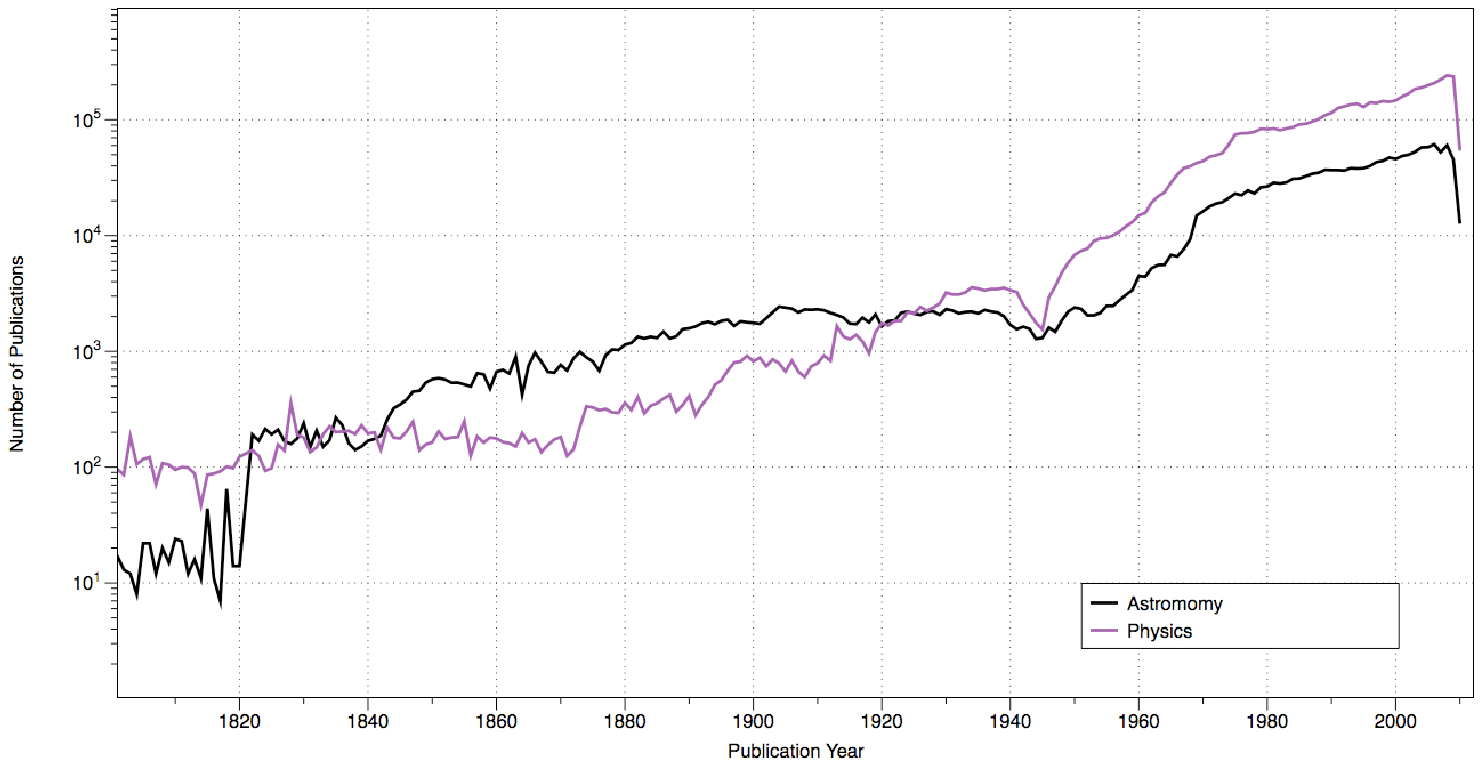}
\includegraphics[scale=0.75]{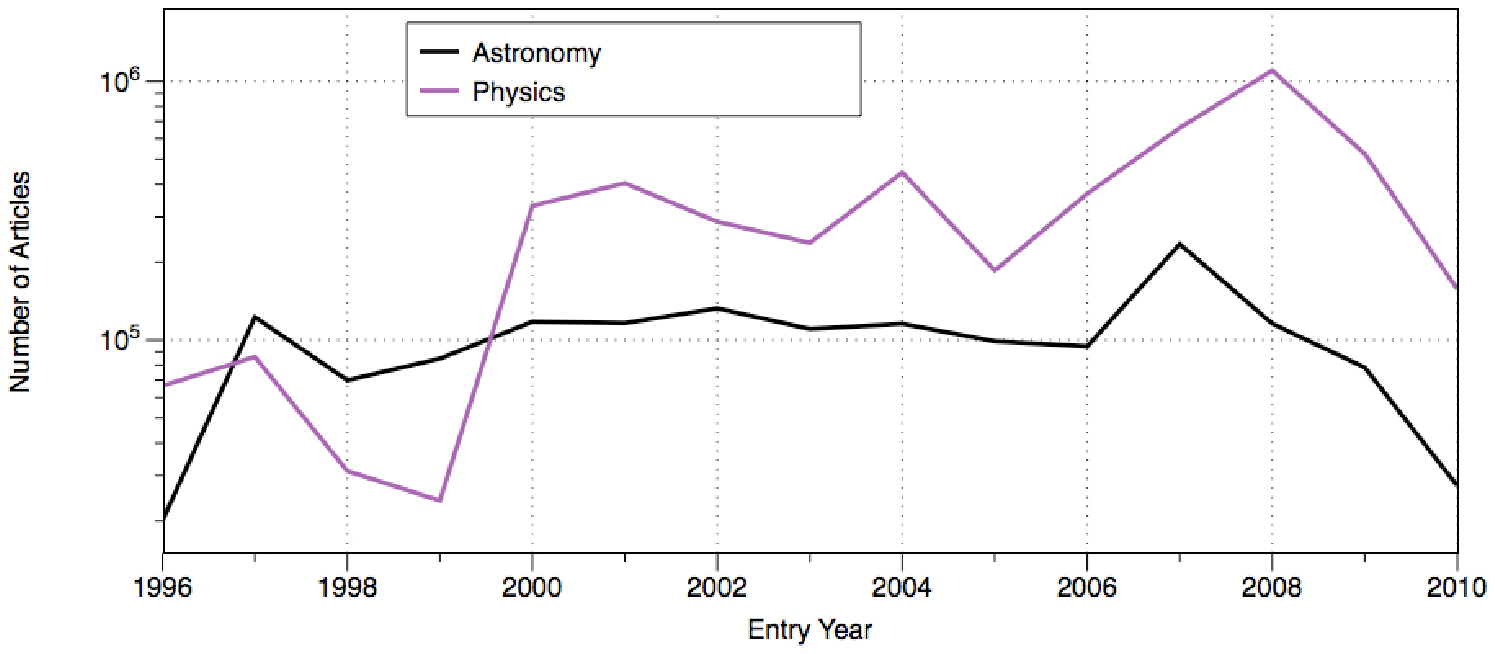}
\caption{Top: Records in the ADS as function of publication year. Bottom: Number of records entered in the ADS}
\label{ADSpubs}
\end{figure}

\begin{figure} [hb]
\centering
\includegraphics[scale=0.75]{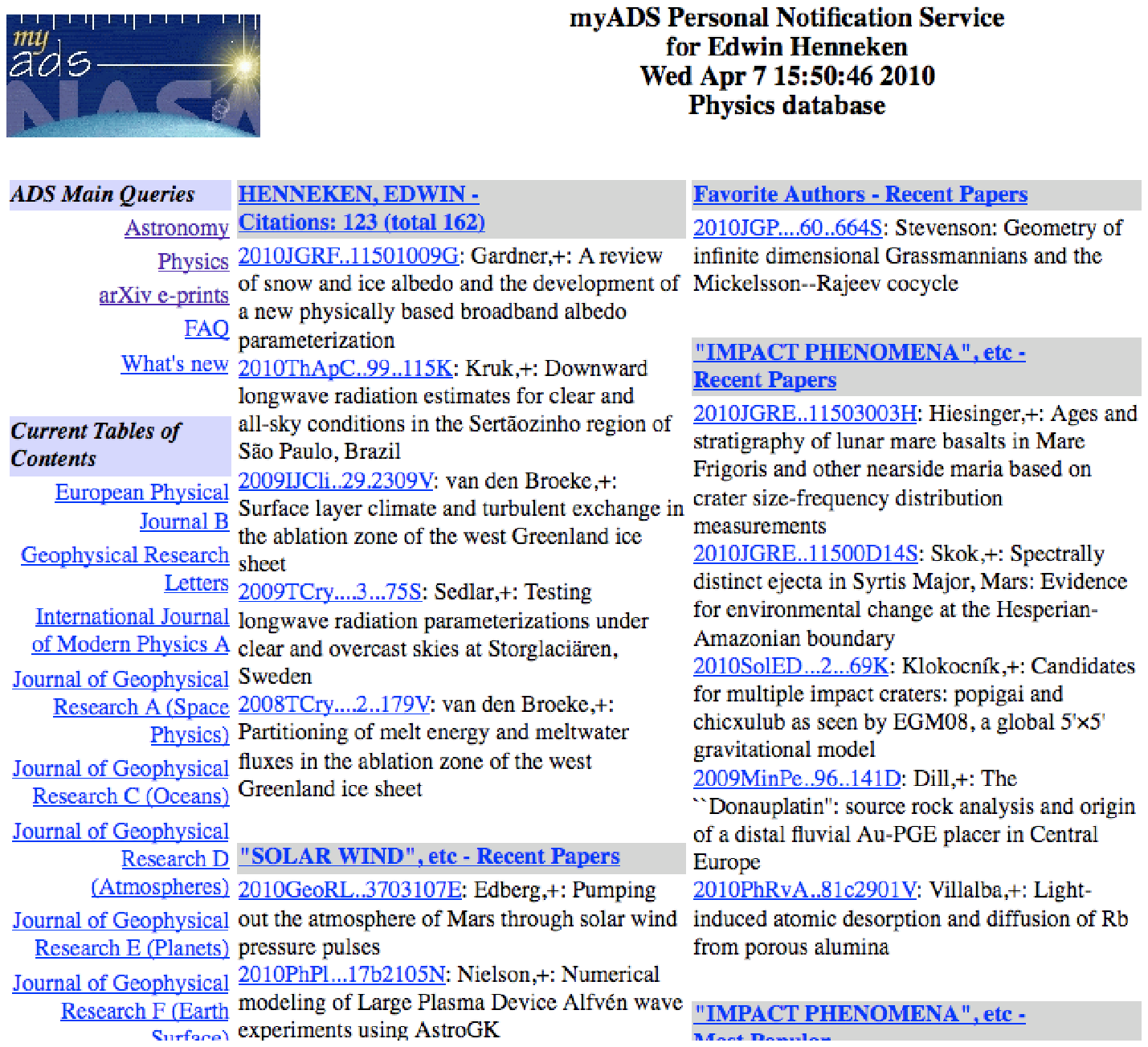}
\caption{The myADS weekly newsletter}
\label{myADS}
\end{figure}

Authors publish because they want to transfer information. An essential ingredient for this transfer is being able to find this information. Of course, it is also in the interest of a publisher or somebody running a search engine service to have efficient and intelligent tools for information discovery. As our Literature Universe is expanding rapidly, finding your way in this deluge of information can be a daunting task. Looking just at the SAO/NASA Astrophysics Data System (ADS) holdings as of April 19, 2010, there are 1,730,210 records in the astronomy database, and 5,437,973 in the physics database, distributed over publication years as shown in figure \ref{ADSpubs} (top). In astronomy, as in other fields, the Literature Universe expands more rapidly because of dissolving boundaries with other fields. Astronomers are publishing in journals and citing articles from journals that had little or no astronomy content not too long ago. Figure \ref{ADSpubs} (bottom) shows how many records were entered in the ADS during the period 1996 - 2010, in both the astronomy and physics database. About 100,000 astronomy records have been added each year, since 2000, with the exception of 2007 when ADS received a large amount of meta data from Springer and ARI (the Astronomisches Rechen-Insitut in Heidelberg, Germany). The large spike for physics in 2008 is due to the ingest of CrossRef meta data.
How do you find what you are looking for and more importantly, information you could not have found using the normal information discovery model? When you have some prior information (like author names and/or subject keywords), you can use your favorite search engine and apply that information as filters. There are also more sophisticated services like myADS (as part of your ADS account, see figure \ref{myADS} for an example and e.g. \cite{henneken07} for more information), that do intelligent filtering for you and provide you with customized suggestions. Alternatively, you can ask somebody you consider to be an expert. This aspect emphasizes that ``finding'' essentially is a bi-directional process. Would it not be nice to have an electronic process that tries to mimic this type of discovery? It is exactly this type of information discovery that recommender systems have been designed for.

\section{Recommender Systems}
Recommender systems can be characterized in the following way. Recommender systems for literature recommendation...
\begin{itemize}
\item are a technological proxy for a social process
\item are a way of suggesting like or similar articles to a user-specific way of thinking
\item try to automate aspects of a completely different information discovery model where people try to find other people considered to be experts and ask them to suggest related articles	
\end{itemize}
In other words, the main goal of a literature recommender system is to help visitors find information (in the form of articles) that was previously unknown to them.

What are the key elements needed to build such a recommender system? The most important ingredient is a ``proximity concept''. You want to be able to say that two articles are related because they are ``closer together'' than articles that are less similar. You also want to be able to say that an article is of interest to a person because of its proximity to that person. The following approach will allow you to do just that:
\begin{itemize}
\item build a ``space'' in which documents and persons can be placed
\item determine a document clustering within this space (``thematic map'')	
\end{itemize}
How do we build such a space? Assigning labels to documents will allow us to associate a ``topic vector'' with each document. This will allow us to assign labels to persons as well (``interest vector''), using the documents they read. Placing persons in this document space can be used in essentially two different ways: use this information to provide personalized recommendations or use usage patterns (``reads'') of expert users as proxies for making recommendations to other users (``collaborative filtering''). As far as the labels themselves are concerned, there are various sources we can distill them from. The most straightforward approach is to use keywords for these labels. One drawback that comes to mind immediately, is the fact that there are no keywords available for historical literature. However, keywords are an excellent labeling agent for current and recent literature.

\begin{figure} [hb]
\centering
\includegraphics[scale=0.75]{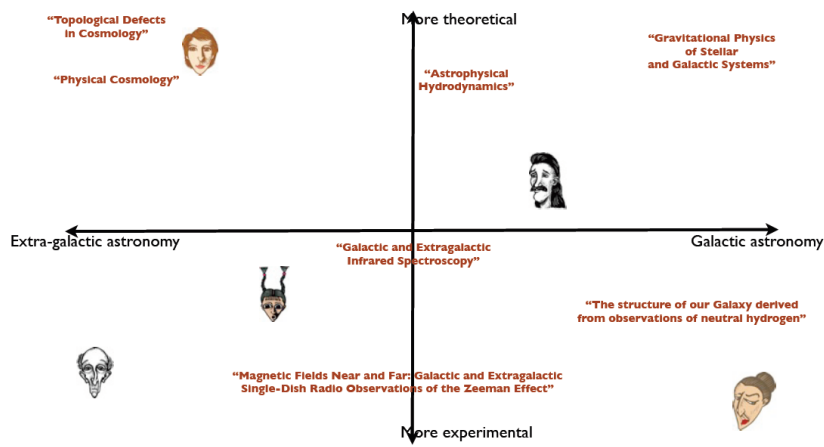}
\caption{A simplistic representation of topic space}
\label{topicspace}
\end{figure}

Figure \ref{topicspace} shows a highly simplified representation of that document space, but it explains the general idea. Imagine a two-dimensional space where one axis represents a topic ranging from galactic to extra-galactic astronomy, and where the other ranges from experimental/observational to theoretical. In this space, a paper titled ``Gravitational Physics of Stellar and Galactic Systems'' would get placed towards the upper right because its content is mostly about theory, with an emphasis on galactic astronomy. A paper titled ``Topological Defects in Cosmology'' would end up towards the upper left, because it is purely theoretical and about the extra-galactic astronomy. A person working in the field of observational/experimental extra-galactic astronomy will most likely read mostly papers related to this subject, and therefore get placed in the lower left region of this space. A clustering is a document grouping that is super-imposed upon this space, which groups together documents that are about similar subjects. As a result, this clustering defines a ``thematic map''. As mentioned, this is a highly simplified example. In reality the space has many dimensions (100 to 200), and these cannot be named as intuitively as ``level of theoretical content''. However, the naming of various directions in this ``topic space'' is not something we need to worry about. The document clustering is the tool that we will be working with.
Now that we have established this ``topic space'', together with the document clustering (defining the ``thematic map''), how do we use this information to find recommendations?

Knowing to which cluster a new article has been assigned will allow us to find papers that are the closest to this article within the cluster. The first couple of papers in this list can be used as a first recommendation. The more interesting recommendations, however, arise when we combine the information we have about the cluster with usage information. The body of usage information is rather specific: it consists of usage information for ``frequent visitors''. People who read between 80 and 300 articles in a period of 6 months seems like a reasonable definition for the group of ``frequent visitors''. We assume that this group of frequent visitors represents either professional scientists or people active in the field in another capacity. People who visit less frequently are not good proxies because they are most likely incidental readers.

To find recommendations from usage data, we determine a group of papers, say 40, that are the closest to the new article within the cluster to which it was assigned. On the basis of this knowledge, we can recommend the following papers:
\begin{enumerate}
\item the paper read the most directly before reading a paper from that group of 40
\item the paper read the most directly after reading a paper from that group of 40
\item the most read paper by the persons who read a paper from the group of 40
\item the most recent paper in the top 100 of the most also-read list from point 3
\item the paper which cites the most papers in that group of 40
\end{enumerate}
where the last recommendation includes citation statistics in the recommendation.
There are more recommendations than the ones mentioned above, but these 5 give a good impression of what is possible.

\section{Example}
As an example, take a recent e-print from arXiv, which, besides the arXiv subject headings, does not come with keywords. Assuming that the bibliography of the e-print contains papers with keywords, we use its bibliography to place the e-print in the document space. Once we know its position in this space, we can establish to which cluster this document belongs.

Michael Kurtz (\cite{kurtz09}) shows an example of the recommender system applied to astronomy.  The creation of the topic space for astronomy included an additional step of creating a “normalized keyword set”, because the set of documents used to create the topic space did not share a unique keyword system. The “normalized keyword set” is essentially a set of identifiers to which the keywords are mapped, using a number of filters and translation rules. Here we want to show an example taken from the physics literature. It was more straightforward to create a topic space for physics because of a large body of documents sharing a unique keyword system: the subject codes from the Physics and Astronomy Classification System (PACS). The topic space was constructed from articles that appeared in the {\it Physical Review} (A,B,C,D,E and Letters), {\it Physics Letters} (A,B), {\it Reviews of Modern Physics}, {\it Nuclear Physics} (A,B), the {\it Journal of Chemical Physics} and {\it Applied Physics Letters}ß, during the period 1995 through 2009. The resulting data set consisted of 144,110 articles. These articles were clustered into 100 clusters containing about 1,000 articles each, on average, using the clustering toolkit ``CLUTO'' (\cite{cluto}) and the statistical toolkit ``R'' (\cite{R}). For usage data we used the anonymized arXiv usage logs for the period of July through December of 2009. 

\begin{figure} [ht]
\centering
\includegraphics[scale=0.75]{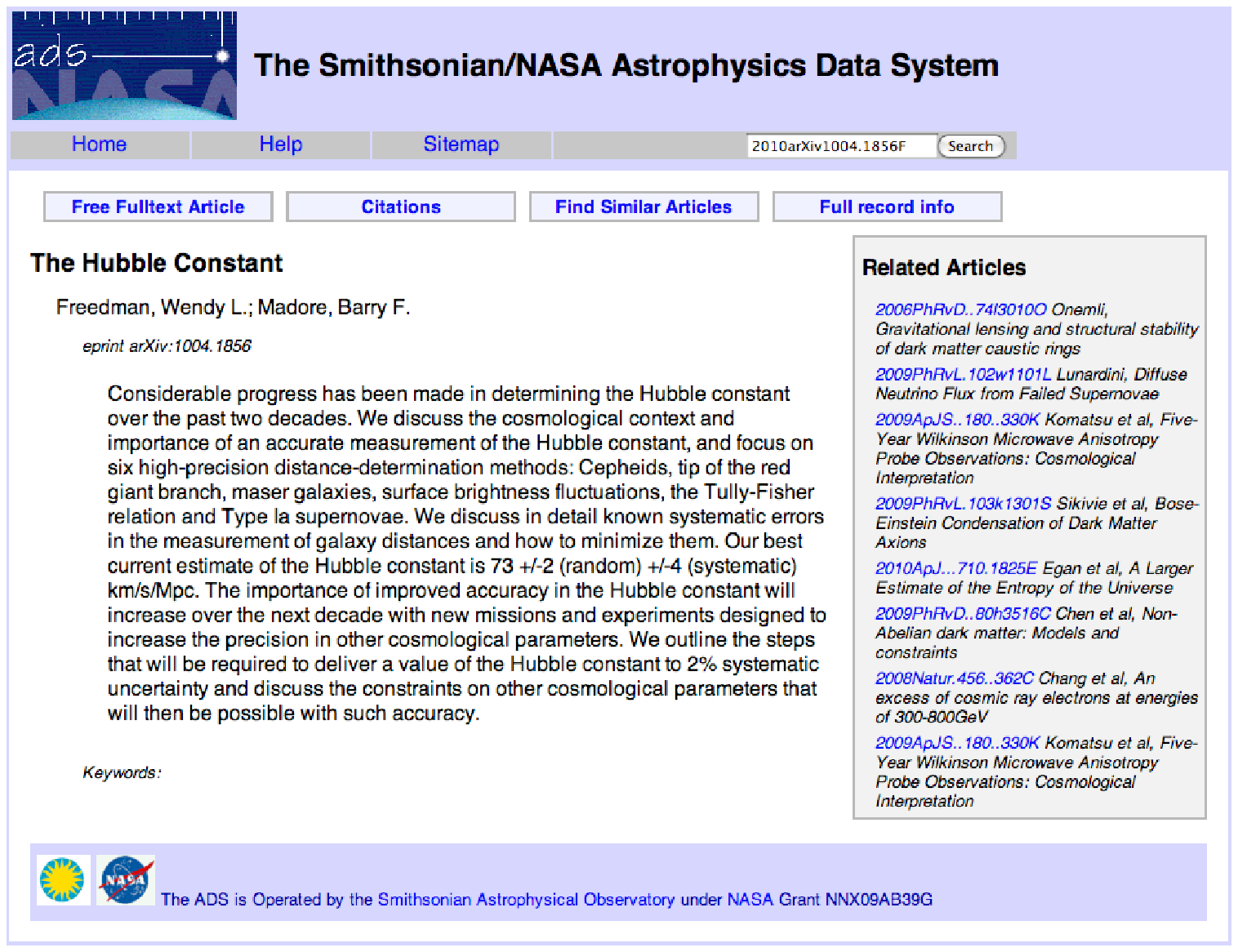}
\caption{Example of Literature Recommendation. The section of ``Related Articles'' has been constructed using PACS identifiers and anonymized usage logs.}
\label{example}
\end{figure}

The example below takes a recent e-print (arXiv:1004.1856), titled ``The Hubble Constant'' by Wendy Freedman and Barry Madore. The bibliography of this e-print allowed us to assign it to a cluster. Based on this cluster assignment and the usage data (see above), the overview of ``Related Articles'' was constructed as shown in figure \ref{example}. All related articles deal with the structure of the Universe and aspects that are related to it, both observational and theoretical. Just from its title ``An excess of cosmic ray electrons at energies of 300-800 GeV'', you might wonder why the Nature article (\cite{chang08}) is among the recommendations. But at closer inspection we see that, among others, the following subjects are discussed in its contents: supernova remnants, dark matter particles and the Wilkinson Microwave Anisotropy Probe (WMAP). These are highly relevant to cosmology in general and the structure of the Universe is particular (see e.g. \cite{spergel2003}). The authors conclude ``If the Kaluza-Klein annihilation explanation proves to be correct, this will necessitate a fuller investigation of such multi-dimensional spaces, with potentially important implications for our understanding of the Universe'', a cosmological statement you might not have expected by just looking at the article title. In this sense the recommender system is not unlike an oracle from Ancient Greece, known for their often cryptic answers. The point we are making here, is the following: the title of an article, and even its abstract, only contain a superficial representation of the article content. Indexing the full article text will obviously result in the best representation of its content, but at the cost of significantly more processing and complexity. One would assume that the key aspects of an article are accompanied with citations of relevant literature, which should make the article bibliography an accurate representation of its content. If keywords classify articles with sufficient accuracy, keywords should therefore be sufficient to base a recommender system on. This illustrates another point with respect to the Nature article we just discussed: because our example was based on the physics literature, using PACS identifiers, all articles from e.g. {\it The Astrophysical Journal} in the bibliography did not participate in the analysis, because they use a different keyword system. Therefore, a recommender system should use labels, in our case keywords, that represent the whole body of literature it is intended for. This means for fields like cosmology and cosmic ray physics, for example, that our recommender system should use both the PACS identifiers and the normalized keyword set, mentioned earlier, in conjunction for optimal accuracy.
\section{Concluding Remarks and Discussion}
The technique used to build the recommender system has been around for quite a while. As early as 1934, Louis Thurstone wrote his paper ``Vectors of the Mind'' (\cite{thurstone34}]) which addressed the problem of ``classifying the temperaments and personality types''. In this paper he extended and generalized Spearman’s method of factor analysis into what is called the ``Centroid Method'' and which became the basis for modern factor analysis. Peter Ossorio (1965) used and built on this technique to develop what he called a ``Classification Space'', which he characterized as ``a Euclidean model for mapping subject matter similarity within a given subject matter domain'' (\cite{ossorio65}). Michael Kurtz (\cite{kurtz93}]) applied this ``Classification Space'' technique to obtain a new type of search method. Where the construction of the ``Classification Space'' in the application by Ossorio relied on data input by human subject matter experts, the method proposed by Michael Kurtz builds the space from a set of classified data. Our recommender system is a direct extension of the ``Statistical Factor Space'' described in the appendix ``Statistical Factor Spaces in the Astrophysical Data System'' of this paper by Michael Kurtz.

We still need to answer the following questions: How stable is the recommender system? What is the best way to update the system? Do we want to allow cluster overlap when creating the thematic map in the topic space? Do keywords really describe the document universe with sufficient accuracy? What do we do with the body of literature for which no keywords are available? Can we optimize the system so that we can provide recommendations ``real-time''?

In order to analyze the stability of the recommender system, we need to explore the dependence of its results on assumptions and simplifications that lie at the basis of the system. For example, we clustered the topic space into 100 clusters of about 1,000 articles on average per cluster. The number of papers (40) we selected for also-read analysis was arbitrary. If we vary these numbers, the results will undoubtedly vary. The main question is: do the results stay relevant?

The Literature Universe keeps expanding, and therefore the topic space needs to be updated on a regular basis. Do we have to recalculate it from scratch, every time, or is there a smart way to update it incrementally? It is probably a bit of both. 

We probably should allow articles to be members of multiple clusters. For example, there are papers that are interesting to some people because they discuss e.g. an observational technique or instrument, and interesting to others because they provide information on a particular astronomical object.

Whether keywords really describe the document universe with sufficient accuracy is directly related to the question whether a keyword system is sufficiently detailed to classify articles. We assume the latter is true. Having said this, we do realize that a keyword system can never be static because of developments within a field and because of diffusing boundaries with other fields. We use the keywords provided by the publishers, so the scope and the evolution of the keyword spectrum is out of our hands. It also means that a recommender system based on publisher-provided keywords has one obvious vulnerability: if a major publisher would decide to stop using keywords (e.g. PACS identifiers), it would pose a significant problem.

One way to build a recommender system without the use of keywords is to use Latent Semantic Indexing. This would also take are of the body of literature for which we do not have keywords, for example historical literature. It is probably not sufficient to index just article abstracts, because of the reason we mentioned in the previous section. However, if we were to index the abstract of an article together with the abstracts of the articles in its bibliography, the level of accuracy would probably be acceptable. This approach is more noisy, but there is enough redundancy in the data to find correlations of a strength similar to the ones we find when using keywords.

The part of the recommender system that needs the most computing power can be pre-computed: the topic space and the resulting clustering. The ``real-time'' part consists of assign a document to a cluster and essentially intersecting a subset of articles from that cluster with usage data. These data sets are small enough to be processed ``real-time'', so we feel that providing visitors with recommendations ``on the fly'' is an achievable goal.

We conclude that a recommender system is feasible for any information retrieval system containing sufficient meta data to classify its records and with a large enough set of regularly returning visitors, providing either implicit (``click stream'') or explicit feedback. Prototype results seem to indicate that using a keyword system for article labeling provides a meaningful literature recommender system. We realize that a different approach will be necessary to include records without keywords.

\acknowledgments
The ADS is Operated by the Smithsonian Astrophysical Observatory under NASA Grant NNX09AB39G. The ADS gratefully acknowledges the support of the arXiv team.

\end{document}